\documentclass[aps,preprint,amsmath,amssymb,showpacs]{revtex4}

\usepackage{graphicx}
\usepackage{color}
\begin{document}
\title{Anomalous Wtb couplings in $\gamma p$ collision at the LHC}

\author{B. \c{S}ahin}
\email[]{bsahin@karaelmas.edu.tr} \affiliation{Department of
Physics, Bulent Ecevit University, 67100 Zonguldak, Turkey}
\author{A. A. Billur}
\email[]{abillur@cumhuriyet.edu.tr} \affiliation{Department of
Physics, Cumhuriyet University, 58140 Sivas, Turkey}

\begin{abstract}

We study the possibility for the process  $pp \rightarrow p\gamma p
\rightarrow pW^-t(W^+\bar t)X$ with anomalous Wtb couplings in a
model independent effective Lagrangian approach at the LHC. We find
$95 \% $ confidence level bounds on the anomalous coupling
parameters for various values of the integrated luminosity. The
improved constraints on the anomalous Wtb couplings have been
obtained compared to current limits.

\end{abstract}


\maketitle

\section{Introduction}

The Standard Model (SM) has been very successful in explaining the
data taken from former colliders such as CERN LEP or Fermilab
Tevatron. Testing SM at the CERN LHC will either lead to additional
confirmation of the SM or give some hints for new physics beyond the
SM. Because of the large mass of the top quark, its couplings are
expected to be more sensitive to new physics than other particles
\cite{peccei,peris}. Especially, Wtb vertex deserves special
attention since the top quark is expected to decay almost completely
via this interaction. Thus, studying top quark couplings will be
substantial to test the SM and a deviation of the top couplings from
the expected values would imply the existence of new physics.

\begin{figure}
\includegraphics{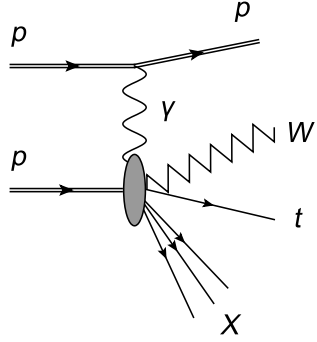}
\caption{Schematic diagram of the process $pp \to p\gamma p \to
pWtX$. \label{fig1}}
\end{figure}

In this work we have analyzed anomalous Wtb couplings via single top
quark production in $\gamma p$ collision at the LHC. This reaction
is probable at the LHC via elastic photon emission from one of the
incoming protons. The emitted photon can collide with the other
proton and produce a final state of $WtX$ through deep inelastic
scattering (Fig.\ref {fig1}). We employ the equivalent photon
approximation (EPA) \cite{Piotrzkowski,Budnev,Baur} for elastic
photon emission from the proton. In the EPA, emitted photons have a
low virtuality and it is a good approximation to assume that they
are on-mass-shell. For this reason these photons are sometimes
called quasi-real photons. When a proton emits a quasi-real photon
it remains intact and scatters with a very small angle from the beam
pipe. The ATLAS and CMS Collaborations at the LHC, have a program
with very forward detectors. It is aimed to investigate soft and
hard diffraction, low x dynamics with forward jet studies, high
energy photon-induced interactions, large rapidity gaps between
forward jets, and luminosity monitoring
\cite{royon1,boonekamp1,boonekamp2,boonekamp3,
khoze1,khoze2,khoze3,khoze4,khoze5,khoze6,kepka,khoze7,schul,klein,gay,
goncalves,machado,rouby}. These detectors will be located in a
region nearly 220-420 m from the interaction point and they can
detect protons in a continuous range of momentum fraction loss
\cite{royon,albrow}. Momentum fraction loss of the proton is defined
as $ \xi=(\vert\overrightarrow{p}\vert-\vert
{\overrightarrow{p}}^{'}\vert)/\vert\overrightarrow{p}\vert $. Here
$ \overrightarrow{p} $ is the momentum of the incoming proton and $
\overrightarrow{p}^{'} $ is the momentum of the intact scattered
proton. Therefore equipped with very forward detectors, LHC can to
some extend be considered as a high-energy photon-photon or
photon-proton collider.

Photon induced rections have been experimentally observed through $
p\overline{p} \rightarrow \gamma \gamma p\overline{p} \rightarrow
\ell^{+} \ell^{-} p \overline{p} $ processes in hadron-hadron
collisions \cite{pp1,pp2,pp3}, $ ep \rightarrow e X p $ in ep
collisions \cite{ep1,ep2,ep3,ep4,ep5,ep6}, and pair production in AA
collisions \cite{AA1,AA2,AA3,AA4}. These experiments raise interest
on the potential of LHC as a photon-photon and photon-proton
collider and motivate phenomenological works on photon-induced
reactions at the LHC as a probe of new physics
\cite{kepka,khoze7,schul,newphlhc1,newphlhc2,newphlhc3,newphlhc4,
newphlhc5,newphlhc6,newphlhc7,newphlhc8,newphlhc9,newphlhc10,newphlhc11}.

\section{Lagrangian and cross sections}

Anomalous $Wtb$ couplings can be investigated in a model independent
way by means of the effective Lagrangian approach
\cite{Buchmuller,Hagiwara,Gounaris,Gounaris2,Whisnant,Yang,Kane}. We
employ the following effective Lagrangian describing anomalous $Wtb$
couplings:
\begin{eqnarray}
\label{lagrangian}
L=\frac{g_{W}}{\sqrt{2}}\left[W_{\mu}\overline{t}(\gamma^{\mu}F_{1L}P_{-}+\gamma^{\mu}F_{1R}P_{+})b
-\frac{1}{2m_{W}}W_{\mu\nu}\overline{t}\sigma^{\mu\nu}(F_{2L}P_{-}+F_{2R}P_{+})b\right]+h.c.
\end{eqnarray}
where
\begin{eqnarray}
W_{\mu\nu}=D_{\mu}W_{\nu}-D_{\nu}W_{\mu}, \,\,\,\,\,\
D_{\mu}=\partial_{\mu}-ieA_{\mu} \nonumber\\
P_{\mp}=\frac{1}{2}(1\mp\gamma_{5}), \,\,\,\,\
\sigma^{\mu\nu}=\frac{i}{2}(\gamma^{\mu}\gamma^{\nu}-\gamma^{\nu}\gamma^{\mu})
\end{eqnarray}
It should be noted that Lagrangian (\ref{lagrangian}) also give rise
to anomalous $Wtb\gamma$ couplings. In the SM, the (V -A) coupling
$F_{1L}$ corresponds to the Cabibbo-Kobayashi-Maskawa (CKM) matrix
element $V_{tb}$, which is very close to unity and $F_{1R}$,
$F_{2L}$ and $F_{2R}$ are equal to zero.

For off-shell top and/or bottom quarks the Lagrangian in
Eq.(\ref{lagrangian}) is not the most general one, it should be
extended with $k^{\mu}$ and $\sigma^{\mu\nu}k_{\nu}$ terms where
$k$ is the sum of the momenta of the t and b quarks. However,
if Wtb couplings arise from gauge invariant effective operators,
single top production and decay can be described in full generality
using the on-shell Lagrangian in Eq.(\ref{lagrangian}) for the Wtb
vertex, even in the process where the top and bottom quarks are far
from their mass shell \cite{AguilarSaavedra:2008gt}. If the W boson
is off-shell, then there are additional terms containing
$\partial_{\mu}W^{\mu}$ \cite{Kane}. These terms are omitted in the
Lagrangian, they can be recovered by applying the equation
of motion through operators of the original Lagrangian \cite{boos2}.

Measurements at D0 detector at Fermilab Tevatron provide stringent
direct constraints on these couplings \cite{d01,d02,d03,d04}. The
most stringent bounds on the anomalous couplings $F_{1R}, F_{2R}$
and $F_{2L}$ are given by $|F_{1R}|^2<0.50, |F_{2R}|^2<0.05$ and
$|F_{2L}|^2<0.11$ at 95\% C.L. assuming that $F_{1L}=1$  \cite{d04}.
Recent results from early LHC data set comparable, but still weaker
bounds with respect to Tevatron \cite{lhc1}. We see from the
Tevatron and LHC data that the bound on the coupling $F_{2L}$ is
weaker then the bound on $F_{2R}$ and the bound on the coupling
$F_{1R}$ is weaker then the others \cite{d01,d02,d03,d04,lhc1}. The
(V +A) coupling $F_{1R}$ is stringently bounded by the CLEO
$b\rightarrow s\gamma$ data \cite {cleo, larios} from an indirect
analysis. Limit from the CLEO data is given by $|F_{1R}|<4\times
10^{-3}$ at $2\sigma$ level\cite{larios}. It is explicit that
indirect constraints are much more resrictive than direct
constraints \cite{drobnak,kolodziej}.

In the literature there has been a great amount of work on Wtb
couplings through single and pair top quark production. The single
top quark production cross section was discussed below and above the
$t\overline{t}$ threshold for the processes $e^{+}e^{-}\rightarrow
Wtb$ \cite {Ambrosanio,Dokholian} and $e^{+}e^{-}\rightarrow
e\overline{\nu}tb$ at the CERN LEP \cite {Hagiwara2, Boos}. Top
quark single and pair production processes were studied for future
linear $e^{+}e^{-}$ collider and its $\gamma\gamma$ and $e\gamma$
modes \cite {Grzadkowski, Grzadkowski2,
Grzadkowski3,Cieckiewicz:2003sd,Kolodziej:2003gp,Batra,Boos:1999ca,Cao:2006pu,sahin}
and also for  $\gamma p$ collisions in TESLA+HERAp and CLIC+LHC
options \cite{atag1,atag2,atag3}. Anomalous Wtb couplings have also
been probed at the LHC and Tevatron \cite
{d01,d02,d03,d04,lhc1,kolodziej,carlson, carlson2, stelzer, Smith,
Hioki, gupta, Berger:2009hi,
AguilarSaavedra:2008gt,Najafabadi:2008pb,AguilarSaavedra:2007rs,delAguila:2002nf}.

The equivalent photon spectrum of virtuality $Q^2$ and energy
$E_\gamma$ is given by the following formula
\cite{Piotrzkowski,Budnev,Baur}
\begin{eqnarray}
\frac{dN_\gamma}{dE_{\gamma}dQ^{2}}=\frac{\alpha}{\pi}\frac{1}{E_{\gamma}Q^{2}}
[(1-\frac{E_{\gamma}}{E})
(1-\frac{Q^{2}_{min}}{Q^{2}})F_{E}+\frac{E^{2}_{\gamma}}{2E^{2}}F_{M}]
\end{eqnarray}
where
\begin{eqnarray}
&&Q^{2}_{min}=\frac{m^{2}_{p}E^{2}_{\gamma}}{E(E-E_{\gamma})},
\;\;\;\;\;\; F_{E}=\frac{4m^{2}_{p}G^{2}_{E}+Q^{2}G^{2}_{M}}
{4m^{2}_{p}+Q^{2}} \nonumber \\
G^{2}_{E}=&&\frac{G^{2}_{M}}{\mu^{2}_{p}}=(1+\frac{Q^{2}}{Q^{2}_{0}})^{-4},
\;\;\;\;\;\;\; F_{M}=G^{2}_{M}  \nonumber
\end{eqnarray}
In Eq.(3), E is the energy of the incoming proton beam and $m_{p}$
is mass of the proton. The magnetic moment of the proton is taken to
be $\mu_{p}^{2}=7.78$ and $Q_{0}^{2}=0.71 GeV^2$
\cite{Piotrzkowski,kepka,rouby}. The photon spectrum which is
integrated from a kinematic minimum $Q^{2}_{min}$ up to
$Q^{2}_{max}$  is given by \cite{kepka}

\begin{eqnarray}
dN(E_{\gamma})=\frac{\alpha}{\pi}\frac{dE_{\gamma}}{E_{\gamma}}\left(1-\frac{E_{\gamma}}{E}\right)
\left[\varphi\left(\frac{Q^{2}_{max}}{Q^{2}_{0}}\right)-\varphi\left(\frac{Q^{2}_{min}}{Q^{2}_{0}}\right)\right]
\end{eqnarray}
here the function $\varphi$ is defined as follows
\begin{eqnarray}
\varphi(x)=(1+ay)\left[-ln(1+x^{-1})+\sum^{3}_{k=1}\frac{1}{k(1+x)^{k}})\right]
+\frac{(1-b)y}{4x(1+x)^{3}}+c\left(1+\frac{y}{4}\right)
\nonumber
\\\times\left[ln\frac{1+x-b}{1+x}+\sum^{3}_{k=1}
\frac{b^{k}}{k(1+x)^{k}}\right]
\end{eqnarray}
where
\begin{eqnarray}
y=\frac{E^{2}_{\gamma}}{E(E-E_{\gamma})},\,\,\,\,\
\nonumber
a=\frac{1}{4}(1+\mu_{p}^{2})+\frac{4m_{p}^{2}}{Q_{0}^{2}}\approx7.16
\end{eqnarray}

\begin{eqnarray}
b=1-\frac{4m_{p}^{2}}{Q_{0}^{2}}\approx-3.96,\,\,\,\,\
c=\frac{\mu_{p}^{2}-1}{b^{4}}\approx0.028
\end{eqnarray}
In the EPA emitted photons have a low virtuality and photon spectrum
has a asymptotic behavior for large values of virtuality $Q^2$.  In
the EPA that we have considered typical photon virtuality is
$\langle Q^2\rangle \approx0.01 GeV^2$ \cite{Piotrzkowski}. Above
this average virtuality value, spectrum function rapidly decreases
and the contribution to the integral above
$Q^{2}_{max}\approx2\;GeV^2$ is negligible. To be precise, the
difference between SM cross sections for $Q_{max}^2=2GeV^2$ and
$Q_{max}^2=64GeV^2$ is at the order of $10^{-5}$pb. Therefore during
calculations we set $Q^{2}_{max}=2\;GeV^2$.

We consider the subprocesses $\gamma b\rightarrow W^{-}t$ and
$\gamma \bar b\rightarrow W^{+}\bar t$ of our main process $pp
\rightarrow p\gamma p \rightarrow pW^-t(W^+\bar t)X$. In the SM
single production of the top quark via the process $\gamma b \to
W^{-}t$ is described by three tree level diagrams. Each of the
diagrams contains a Wtb vertex. In the effective Lagrangian
approach, there are four tree level diagrams; one of them contains
an anomalous $\gamma btW$ vertex, which is absent in the SM (Fig.2).

The total cross section for the process $pp \rightarrow p\gamma p
\rightarrow pW^-t(W^+\bar t)X$ can be obtained by integrating the
cross section for the subprocesses over the photon and quark
distributions:
\begin{eqnarray}
\sigma\left(pp \rightarrow p\gamma p \rightarrow pW^-t(W^+\bar
t)X\right)=&&\int_{\xi_{1\; min}}^{\xi_{1\;max}} {dx_1 }\int_{0}^{1}
{dx_2}
\left(\frac{dN_\gamma}{dx_1}\right)\left(\frac{dN_q}{dx_2}\right)\nonumber
\\&&\times \left[\hat{\sigma}_{\gamma b\rightarrow W^{-}t}(\hat
s)+\hat{\sigma}_{\gamma \bar b\rightarrow W^{+}\bar t}(\hat
s)\right]
\end{eqnarray}
In this formula, $x_{1}=\frac{E_{\gamma}}{E}$ and $x_{2}$ is the
fraction which represents the ratio between $b$ $(\bar b)$ quark and
incoming proton's momentum. $\frac {dN_{q}} {dx_{2}}$ is the $b$
$(\bar b)$ quark distribution function. We ignore interactions
between different family quarks since the cross sections are
suppressed due to small off diagonal elements of the
Cabibbo-Kobayashi-Maskawa matrix.  In the total cross section
calculations we have used Martin, Stirling, Thorne and Watt
distribution functions \cite{martin}.

In our calculations three different forward detector acceptance
ranges have been discussed: $ 0.0015<\xi<0.15 $, $ 0.0015<\xi<0.5$
and $ 0.1<\xi<0.5 $. The former one was proposed by the ATLAS Forward Physics
(AFP) Collaboration \cite{royon, albrow}. The second acceptance
range was proposed by the CMS-TOTEM forward detector scenario \cite{cms-totem}.
Since the forward detectors can detect protons
in a continuous range of $\xi$ one can impose some cuts and choose to work in a
subinterval of the whole acceptance region. Hence, we also consider an acceptance
of $0.1<\xi<0.5 $ which is a subinterval of the CMS-TOTEM acceptance
range.

In Figs. 3-6 we show the integrated total cross
section of the process $pp \rightarrow p \gamma p \rightarrow pW^{-}tX$ as
a function of anomalous couplings $ F_{2R} $, $ F_{2L}$, $F_{1R}$
and $\Delta F_{1L}$  for the
acceptances of $ 0.0015<\xi<0.15 $, $ 0.0015<\xi<0.5 $ and $ 0.1<\xi<0.5$. Here, the anomalous
coupling $\Delta F_{1L}$ is defined by $\Delta F_{1L}\equiv F_{1L}-0.99$. In Figs. 3 and 4 we observe that
cross section approximately has a same dependence to both $F_{2R} $ and $ F_{2L}$ couplings. We see from Fig. 5
that sensitivity of the cross section to anomalous coupling $F_{1R}$ is comparably weak.

\section{Sensitivity to anomalous couplings}

We estimate the sensitivity of the process $pp \rightarrow p\gamma p \rightarrow pW^-t(W^+\bar
t)X$ to anomalous couplings $ F_{2R} $, $ F_{2L}$, $F_{1R}$ and $\Delta F_{1L}$ using a simple one parameter
$\chi^{2}$ criterion for integrated luminosities of $L_{int}=10, 30, 50, 100, 200 fb^{-1}$ and $\sqrt{s}$ = 14 TeV.
The $\chi^{2}$ function is given by
\begin{eqnarray}
\chi^{2}=\left(\frac{\sigma_{SM}-\sigma(\Delta
F_{1L},F_{1R},F_{2L},F_{2R})}{\sigma_{SM}\delta}\right)^{2}
\end{eqnarray}
where $\delta=\frac{1}{\sqrt{N}}$ is the statistical error. The
expected number of events has been calculated considering the
leptonic decay channel of the W boson and leptonic decay of the
top quark as the signal $N=BR(W^{-}\rightarrow \ell^{-}
\overline{\nu}_{\ell})BR(t\rightarrow W^{+}b\rightarrow
\ell^{+}\nu_{\ell}b)\sigma_{SM}L_{int}$, where $\ell=e$ or $\mu$.
ATLAS and CMS have central detectors with a pseudorapidity coverage $\vert\eta\vert<2.5$.
Therefore we consider an acceptance window of $\vert\eta\vert<2.5$ for final state electrons,
 muons and b quark. Branching ratios appearing in the number of events are defined as
$BR=\frac{\Gamma} {\Gamma_{total}}$ where $\Gamma_{total}$ is the
full width and $\Gamma$ is the decay rate for the corresponding
channel with a cut of $|\eta|<2.5$ for final decay products.

The limits for the anomalous coupling parameters are given in tables
1-4 for integrated luminosities of $L_{int}=10, 30, 50, 100, 200
fb^{-1}$ and forward detector acceptances of $ 0.0015<\xi<0.15 $, $
0.0015<\xi<0.5$ and $ 0.1<\xi<0.5 $. We see from the tables that $
0.1<\xi<0.5 $ case provides more restrictive bounds on both $ F_{2R}
$ and $ F_{2L}$ couplings compare to the $ 0.0015<\xi<0.15 $ and $
0.0015<\xi<0.5$ cases. On the other hand, limits on $F_{1R}$ and
$\Delta F_{1L}$ couplings in $0.1<\xi<0.5$ case are weaker then the
limits in $ 0.0015<\xi<0.15 $ and $ 0.0015<\xi<0.5$ cases. The
limits presented in tables are reasonable. In the effective
lagrangian (1) anomalous couplings $F_{1R}$ and $F_{1L}$ are
originated from dimension 4 effective operators but anomalous
couplings $F_{2R}$ and $F_{2L}$ are originated from dimension 5
effective operators. Therefore energy dependence of the terms in the
cross section proportional to $F_{2R}$ and $F_{2L}$ are expected to
be higher than the standard model and also the terms in the cross
section proportional to $F_{1R}$ and $F_{1L}$. Hence, the main new
physics contribution from couplings $F_{2R}$ and $F_{2L}$ comes from
high energy region $0.1<\xi<0.5$. On the contrary, the main standard
model contribution comes from low energy region $0.0015<\xi<0.15$.
The limits on $F_{2R}$ and $F_{2L}$ are expected to be better in the
less forward region $0.1<\xi<0.5$ since the invariant mass of the
incoming photon and b quark is large and the standard model
background is low in that region.

\section{Conclusion}

Our limits on the couplings $F_{2R}$  and $F_{2L}$ are approximately a factor from 2 to 4 and 3 to 6 better then the limits
from direct constraints at the Tevatron respectively depending on the luminosity \cite{d04}. On the other hand, our best
limit on $F_{1R}$ is a factor of 3.7 more restricted compared to Tevatron direct constraint \cite{d04}.
%Experimental bounds on $F_{1L}$ can be extracted from the constraints on CKM matrix element $V_{tb}$.
%The direct determination of $|V_{tb}|$ without any assumption about unitarity implies $|V_{tb}|=0.88 \pm 0.07$ \cite{PDG2010}.

Physics studied at the LHC is significantly enhanced via the forward physics programs of ATLAS and CMS collaborations.
Equipped with forward detectors LHC gives us new options to examine high energy photon-proton interactions. With respect to
pure deep inelastic scattering processes, photon-proton interactions provide a quite clean channel due to absence of one of
the incoming proton remnants. Furthermore, detection of the intact scattered protons in the forward detectors allows us to
reconstruct quasi-real photons momenta. This provides an advantage in reconstruction of the kinematics.

\pagebreak

\begin{figure}
\includegraphics{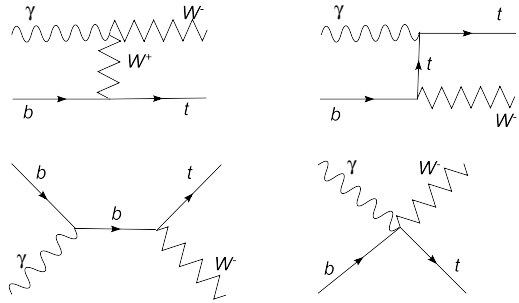}
\caption{Tree level Feynman diagrams for the process $\gamma b \to
W^{-}t$. \label{fig2}}
\end{figure}

\begin{figure}
\includegraphics{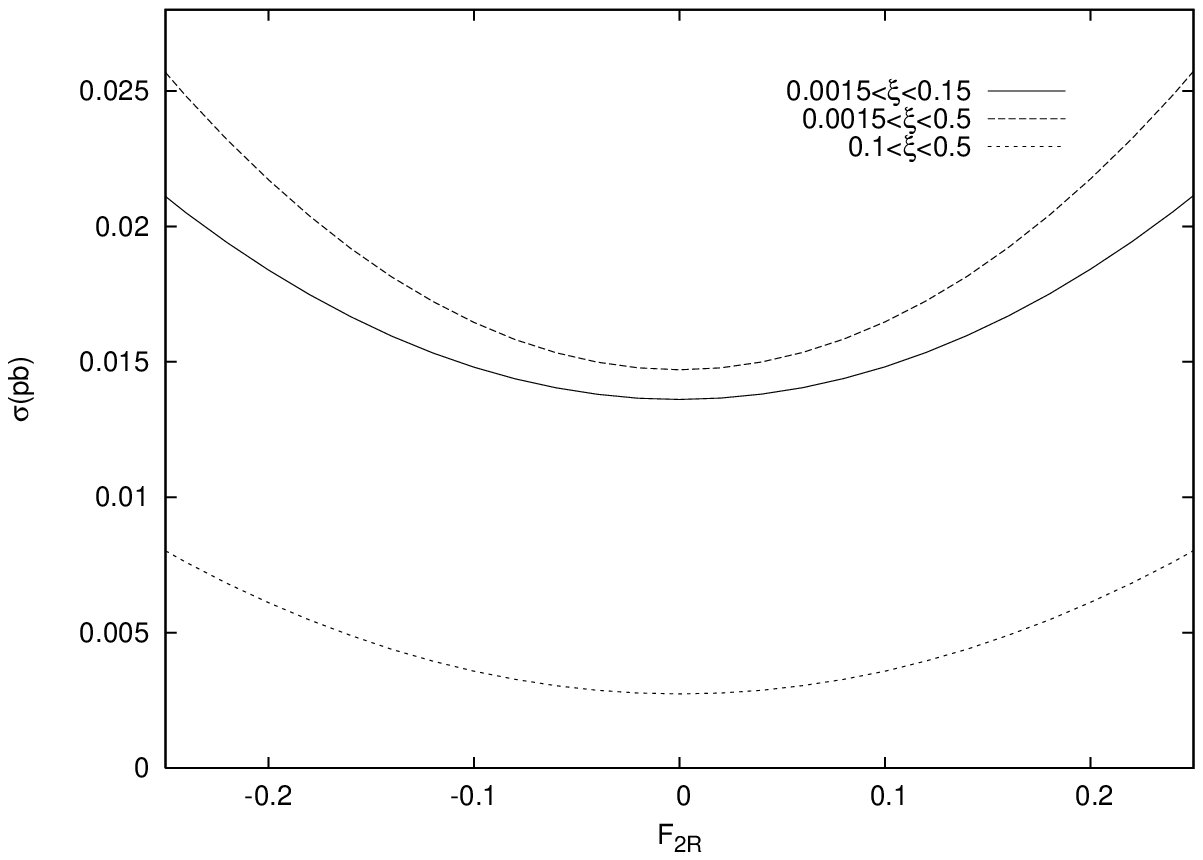}
\caption{ \label{fig3} The integrated total cross section of the
process $pp \to p\gamma p \to pWtX$ as a function of anomalous
coupling $F_{2R}$ for three different detector acceptances stated in
the figure. The center of mass energy is taken to be $ \sqrt{s}=14$
TeV.}
\end{figure}

\begin{figure}
\includegraphics{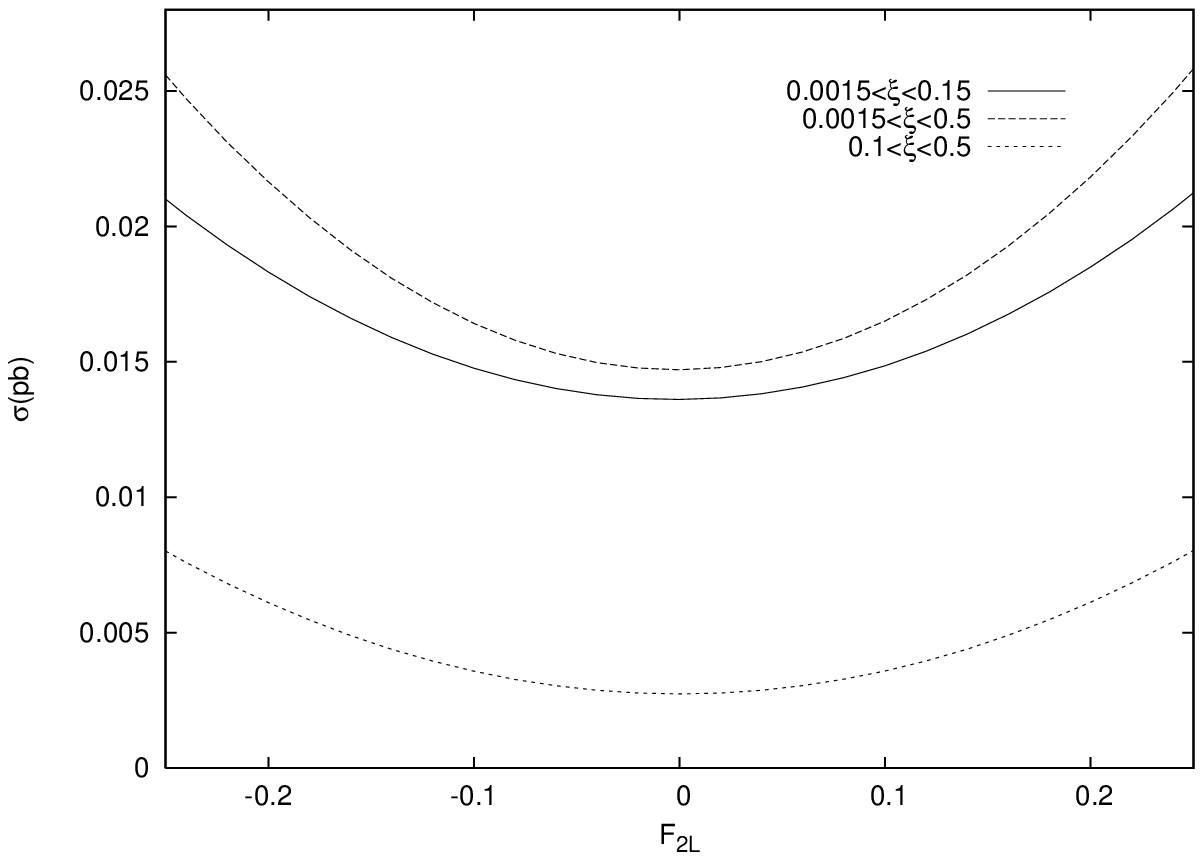}
\caption{ \label{fig4} The integrated total cross section of the
process $pp \to p\gamma p \to pWtX$ as a function of anomalous
coupling $F_{2L}$ for three different detector acceptances stated in
the figure. The center of mass energy is taken to be $ \sqrt{s}=14$
TeV.}
\end{figure}

\begin{figure}
\includegraphics{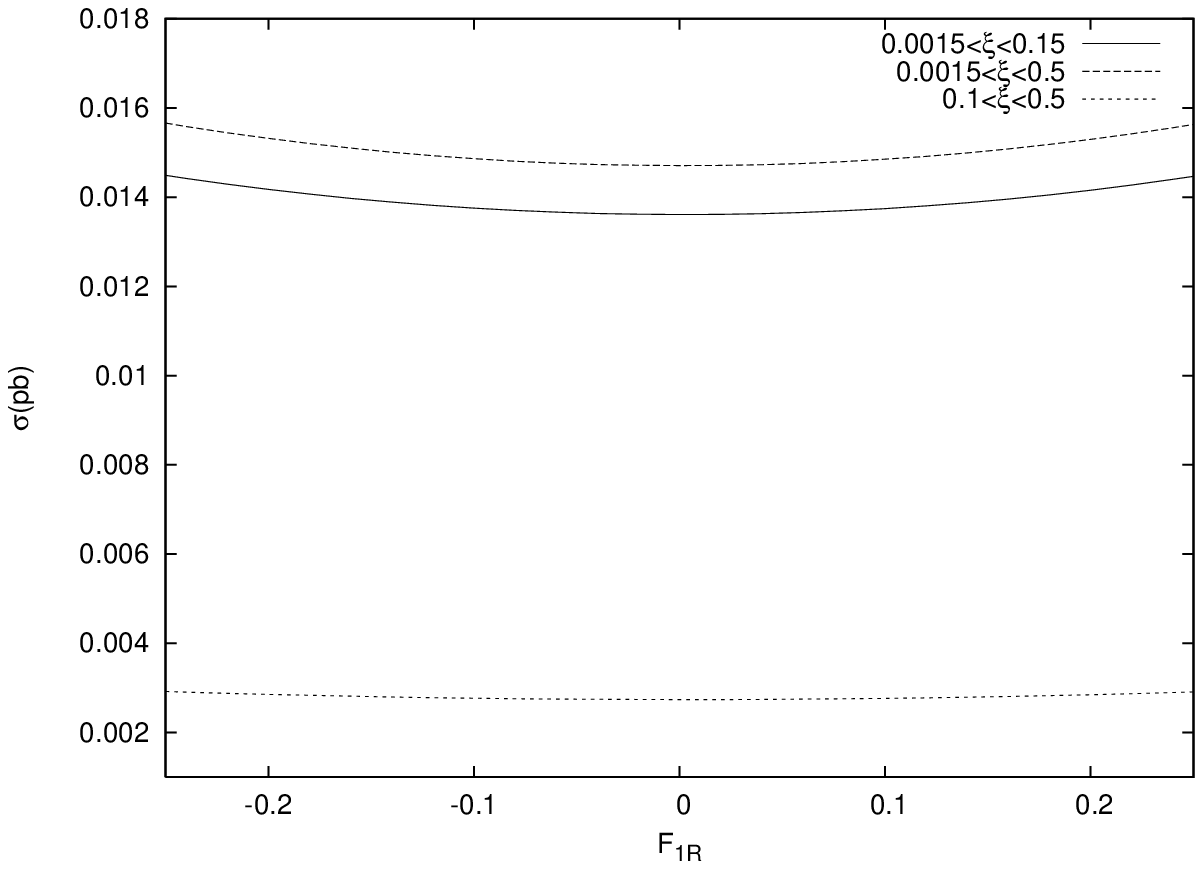}
\caption{ \label{fig5} The integrated total cross section of the
process $pp \to p\gamma p \to pWtX$ as a function of anomalous
coupling $F_{1R}$ for three different detector acceptances stated in
the figure. The center of mass energy is taken to be $ \sqrt{s}=14$
TeV.}
\end{figure}

\begin{figure}
\includegraphics{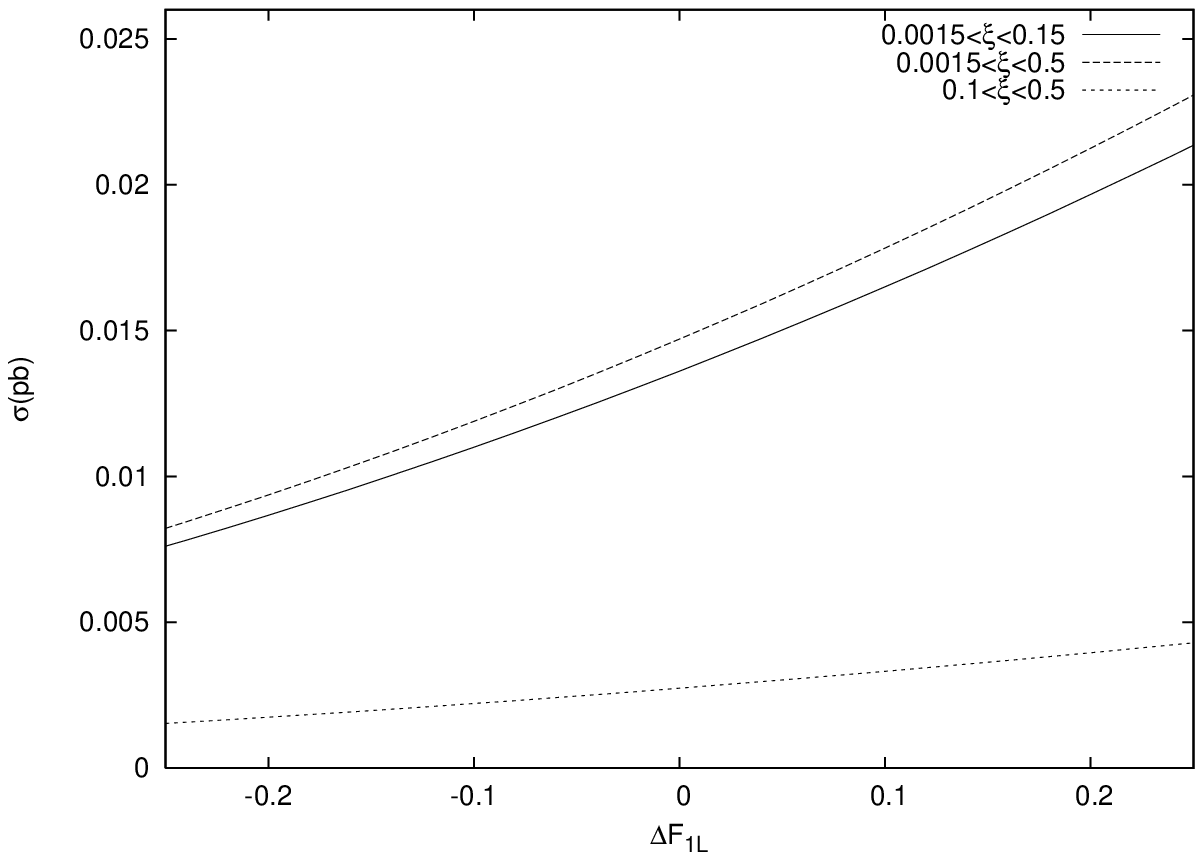}
\caption{ \label{fig6} The integrated total cross section of the
process $pp \to p\gamma p \to pWtX$ as a function of anomalous
coupling $\Delta F_{1L}$ for three different detector acceptances
stated in the figure. The center of mass energy is taken to be $
\sqrt{s}=14$ TeV.}
\end{figure}

\pagebreak

\begin{table}
\caption{95\% C.L. sensitivity bounds of the
coupling $ F_{2R} $ for various forward detector acceptances and
integrated LHC luminosities. The center of mass energy of the
proton-proton system is taken to be $ \sqrt{s}=14 $ TeV.}
\begin{ruledtabular}
\begin{tabular}{ccccccc}
 $L(fb^{-1})$ & $0.0015<\xi<0.5$ & $0.0015<\xi<0.15$ &  $0.1<\xi<0.5$\\
 \hline
  10  &        -0.117;0.116   &   -0.138;0.138  & -0.110;0.110 \\
  30  &        -0.089;0.088   &   -0.105;0.105  & -0.084;0.084 \\
  50  &        -0.078;0.078   &   -0.093;0.092    & -0.074;0.074 \\
  100 &        -0.066;0.065   &   -0.078;0.077  & -0.062;0.062 \\
  200 &        -0.055;0.055   &   -0.066;0.065  & -0.052;0.052\\
\end{tabular}
\end{ruledtabular}
\end{table}

\begin{table}
\caption{95\% C.L. sensitivity bounds of the
coupling $ F_{2L} $ for various forward detector acceptances and
integrated LHC luminosities. The center-of-mass energy of the
proton-proton system is taken to be $ \sqrt{s}=14 $ TeV.}
\begin{ruledtabular}
\begin{tabular}{ccccccc}
$L(fb^{-1})$ & $0.0015<\xi<0.5$ & $0.0015<\xi<0.15$ &  $0.1<\xi<0.5$\\
\hline
  10  &     -0.118;0.115      &  -0.140;0.136   &  -0.110;0.110\\
  30  &     -0.090;0.087      &   -0.107;0.103  &  -0.084;0.084\\
  50  &     -0.079;0.077      &   -0.094;0.090  &  -0.074;0.074\\
  100 &     -0.067;0.064      &   -0.080;0.076  & -0.062;0.062 \\
  200 &     -0.056;0.054      &   -0.067;0.063  & -0.052;0.052\\
\end{tabular}
\end{ruledtabular}
\end{table}

\begin{table}
\caption{95\% C.L. sensitivity bounds of the
coupling $ F_{1R} $ for various forward detector acceptances and
integrated LHC luminosities. The center-of-mass energy of the
proton-proton system is taken to be $ \sqrt{s}=14 $ TeV.}
\begin{ruledtabular}
\begin{tabular}{ccccccc}
 $L(fb^{-1})$ & $0.0015<\xi<0.5$ & $0.0015<\xi<0.15$ &  $0.1<\xi<0.5$\\
 \hline
  10  &        -0.396;0.400   &  -0.404;0.408   & -0.603;0.609 \\
  30  &        -0.300;0.304   &   -0.307;0.310  & -0.457;0.464 \\
  50  &        -0.264;0.268   &    -0.270;0.273 & -0.402;0.409 \\
  100 &        -0.222;0.226   &    -0.227;0.230 & -0.337;0.344 \\
  200 &        -0.186;0.190   &    -0.190;0.194 & -0.283;0.290\\
\end{tabular}
\end{ruledtabular}
\end{table}

\begin{table}
\caption{95\% C.L. sensitivity bounds of the
coupling $\Delta F_{1L} $ for various forward detector acceptances
and integrated LHC luminosities. The center-of-mass energy of the
proton-proton system is taken to be $ \sqrt{s}=14 $ TeV.}
\begin{ruledtabular}
\begin{tabular}{ccccccc}
 $L(fb^{-1})$ & $0.0015<\xi<0.5$ & $0.0015<\xi<0.15$ &  $0.1<\xi<0.5$\\
 \hline
  10  &        -0.084;0.077   &   -0.087;0.080  & -0.207;0.171 \\
  30  &        -0.047;0.045   &    -0.049;0.047 & -0.114;0.102 \\
  50  &        -0.036;0.035   &    -0.038;0.037 & -0.087;0.080 \\
  100 &        -0.026;0.025   &    -0.027;0.026 & -0.060;0.057 \\
  200 &        -0.018;0.018   &    -0.019;0.018 & -0.042;0.041\\
\end{tabular}
\end{ruledtabular}
\end{table}

\end{document}